# A Noachian proglacial paleolake on Mars: Fluvial activity and lake formation within a closed-source drainage basin crater and implications for early Mars climate

Benjamin D. Boatwright[1], James W. Head[1]

ABSTRACT

A 54-km diameter Noachian-aged crater in the southern highlands of Mars contains unusually well-preserved inverted fluvial channel networks and lacustrine deposits, all of which formed completely inside the crater. This "closed-source drainage basin" (CSDB) crater is distinct from previously documented fluvially breached or groundwater-fed crater basin lakes on Mars. We compare our observations to previously established models of crater degradation, fluvial incision, and topographic inversion on Mars to assess the most likely origins of the water that formed the fluvial and lacustrine features. We favor top-down melting of a cold-based glacier as the source of water in the CSDB crater, which would represent the first examples of proglacial fluvial channels and lakes found on Noachian Mars.

---

[1] Department of Earth, Environmental and Planetary Sciences, Brown University, Providence, RI 02912 USA. Corresponding author: Benjamin D. Boatwright, benjamin_boatwright@brown.edu

# 1. INTRODUCTION AND BACKGROUND

Geologic evidence suggests that the ambient climate of early Mars (Noachian; ~4.0—3.6 Ga) was significantly different than today, with compelling evidence for extensive fluvial channels and crater basin lakes, increased erosion rates, and enhanced crater degradation (Craddock & Howard 2002; Howard et al. 2005; Irwin et al. 2005). Indeed, compared to younger craters, most Noachian-aged craters have 1) lowered or completely removed rims; 2) flat, infilled floors; and 3) walls and ejecta deposits modified by fluvial incision and backwasting (Craddock et al. 1997; Craddock & Howard 2002; Mangold et al. 2012). Morphometric studies have interpreted this set of characteristics to indicate degradation by a dual advective—diffusive process (Craddock et al. 1997; Craddock & Howard 2002; Forsberg-Taylor et al. 2004; Matsubara et al. 2018). One such pair of processes involves diffusive rainsplash and advective runoff from rainfall occurring intermittently over tens of millions of years (e.g. Hoke et al. 2011) in a warm and wet Noachian climate (e.g. Craddock & Howard 2002). Although not favored by Craddock and Howard (2002), they also noted that "it is possible that solifluction generated the diffusional component [of crater degradation] while any surface runoff was the result of snowmelt." This would favor a colder, potentially subfreezing Noachian climate with only transient warming. Some global climate modeling studies support ambient Noachian temperatures well below freezing (Forget et al. 2013; Wordsworth et al. 2013, 2015) with cold-based glaciation occurring in the southern highlands (Head & Marchant 2014; Fastook & Head 2015). Thus, the nature of the ambient Noachian martian climate is currently debated.

In addition to their degraded morphology, numerous Noachian-aged craters on Mars have been identified as candidate sites of former lakes. Fassett & Head (2008b) identified 210 open-basin lakes (OBLs), craters fed by valley networks that contained outlet valleys indicative of lake filling and overflow. Goudge et al. (2015) identified an additional 205 candidate closed-basin lakes (CBLs), similar to OBLs but lacking outlet channels. Most OBLs and some CBLs are fed by regionally integrated valley networks, suggesting that these paleolake basins formed at a time similar to the peak of valley network activity around the Noachian—Hesperian boundary (Fassett & Head 2008a), and thus may have required sustained warm and wet climate conditions in order to form (e.g. Irwin et al. 2005; Howard 2007; Matsubara et al. 2011). Salese et al. (2019) identified 24 craters near the dichotomy boundary with floor elevations below −4 km that contained crater wall sapping valleys, terraces, and deltas that they interpreted to be sourced from groundwater upwelling; these also appear to have formed around the Noachian—Hesperian boundary.

We report on the geology of a 54-km diameter Noachian-aged crater in the southern highlands (Figs. 1-3) that contains unusually well-preserved inverted fluvial channel networks and lacustrine deposits. Unlike previously described OBLs and CBLs (Fassett & Head 2008b; Goudge et al. 2015), this crater has neither inlet nor outlet channels and shows no evidence of sapping valleys or other associated landforms suggesting groundwater processes (e.g. Salese et al. 2019). On the basis of these unique characteristics, we term this new type of crater a "closed-source drainage basin" (CSDB), distinct from either fluvially breached or groundwater-fed crater basin lakes on Mars. We compare possible water sources for the fluvial and lacustrine features within the CSDB crater with the proposed origins of other crater basin lakes and inverted terrain features on Mars. We favor top-down melting of a cold-based glacier as the source of water in the CSDB crater. This interpretation would provide observational evidence of the cold and icy early Mars climate predicted by certain models and would represent the first examples of proglacial fluvial channels and lakes found on Noachian Mars.

## 2. GEOLOGY OF THE CSDB CRATER

The 54-km crater (Figs. 1-3; 20.3°S, 42.6°E) is located in Noachian-aged highlands terrain (Tanaka et al. 2014) ~800 km northwest of the Hellas basin rim in Terra Sabaea. In a regional study of Terra Sabaea crater floors, Irwin et al. (2018) noted that the interior of this crater, which they designated "B," contained a dark-over-light stratigraphy consisting of a more resistant lower unit underlying a more erodible upper unit outcropping as knobs and mesas. The lower light-toned unit exhibits an unusual spectral signature consistent with $Fe^{2+}$ substitution in plagioclase (Wray et al. 2013; Irwin et al. 2018) that sets it apart from more prevalent light-toned units in the region such as phyllosilicates and chlorides. Irwin et al. (2018) identified Al and Fe/Mg smectite in the walls of crater B, but these are both morphologically and spectrally distinct from the light-toned floor. Sinuous ridges preserved in the dark-toned upper unit were interpreted by Irwin et al. (2018) as inverted fluvial channels.

On the basis of this previous work, we mapped the rim, walls, and interior of crater B in detail using 5 m/pix Context Camera (CTX, Malin et al. 2007; Fig. 1) and 0.25-0.5 m/pix High Resolution Imaging Science

Experiment (HiRISE, McEwen et al. 2007) visible images; 100 m/pix Thermal Emission Imaging System (THEMIS, Christensen et al. 2004) daytime infrared images; and 25 m/pix CTX stereo topography (Figs. 2-3). Our mapping reveals an unbreached crater rim and an ensemble of closely related geomorphic features in the crater interior (Fig. 1) that we describe below: 1) tributary, linear to sinuous ridges extending downslope from below the base of the crater wall; 2) deflated depositional basins on the crater floor; and 3) a series of crater wall alcoves and upslope-facing scarps above the ridge heads.

## 2.1. Ridges

Tributary ridges typically ~200 m wide and ~20 m high are distributed throughout the crater floor (Fig. 1A, blue); two tributary networks are particularly well preserved in the east (Fig. 1B) and south (Fig. 1C). They both follow the same general morphologic trend, extending radially from below the crater wall base, along the gently sloping floor to the lowest points of the crater (Fig. 2A), where they terminate in now-deflated depositional basins. Their proximal morphology is characterized by bifurcated clusters of narrow, subparallel ridges whose heads begin ~1 km below the upslope-facing scarps. The proximal ridge network in the east merges into a wider, flat-topped medial morphology downslope; approximately a dozen medial ridges then coalesce into four larger trunk ridges (Fig. 1B). The smaller network in the south maintains a mostly medial morphology before narrowing into smaller ridges near their termini; one ridge displays an anabranching morphology (Fig. 1C). We interpret these ridges as networks of fluvial channels that have been preserved in inverted relief, in agreement with Irwin et al. (2018).

## 2.2. Depositional Basins

Two depositional basins are located in the lowest parts of the crater floor (Fig. 1A, cyan/green). The larger basin I is outlined by outcrops of the lower light-toned floor unit. Most of the ridges that originate from below the upslope-facing scarps on the eastern crater wall terminate in basin I. Basin II is readily identified by its distinct texture of features we interpret as transverse aeolian ridges (TARs; e.g. Balme et al. 2008) that are more densely spaced than in the surrounding crater floor.

On the basis of topography and morphology (Figs. 1-2), we interpret these two basins as former lakes and depocenters for sediment transported by the inverted fluvial channels. The mouths of the trunk streams flowing into basin I follow a topographic contour at ~2050 m (Fig. 2D), suggesting they were controlled by an equipotential surface. Channels flowing into basin II terminate within meters of the morphological boundary of the basin in several locations, which could indicate rapid dispersal of sediment into a standing body of water (Fig. 1C).

Irwin et al. (2018) interpreted the two basins as distinct units, but our analyses indicate that the two basins consist of the same bedrock lithology with varying degrees of aeolian cover. This aeolian sediment is likely to have been reworked from the original lacustrine deposits that filled the two basins and subsequently revealed the underlying light-toned bedrock. The concentration of TARs within basin II in consistent with an increased supply of fine-grained sediments potentially derived from the former lakebed. The lack of typical depositional morphologies such as fans and deltas within either basin suggests that the instantaneous sediment supply into the basins was relatively low, but enough sediment accumulated for it to be reworked or removed by subsequent aeolian activity.

We performed order of magnitude calculations of sediment flux through the channels following the paleohydrologic reconstruction methods of Rosenberg et al. (2019) and Hayden et al. (2019), assuming that the ridges represent channel belt deposits as opposed to single channel fills (Appendix B). Assuming only one channel was active at any given time, our calculations indicate that the paleolake basins could be filled with sediment in less than a single Earth year of constant flow. Given the more likely case of intermittent or seasonal activity, the channels were likely active over a period no greater than ~$10^3$ years.

### 2.3. Alcoves and Upslope-Facing Scarps

Subdued alcoves are seen in the steepest sections of the northern and eastern crater wall with average slopes of ~15°. Below the alcoves and approximately 1 km above the proximal ridge heads is a discontinuous scarp consisting of knobby, arcuate segments (Figs. 1A, 2B-C). In some cases, the margins curve upward and extend parallel to the slope, forming a convex-downward U shape. The scarps typically display a steeper upslope face and a shallower downslope face with a relief of ~10-30 m (Figs. 2D, 4D). Analogous features were noted by Davis et al. (2019) in association with inverted fluvial channels in Barth crater (Fig. 4E), but upslope-facing scarps have not otherwise been described in degraded Noachian-aged craters (e.g. Craddock et al. 1997; Craddock & Howard 2002; Irwin et al.

2018). The distribution of the scarps and alcoves is often coincident with clusters of proximal ridge heads, suggesting that all three features may have formed through a common process.

### 2.4. Mantling Deposit and Topographic Inversion

The presence of inverted channels, the abundance of TARs, and the exposures of crater floor substrate material testify to the removal of several tens of meters of sediment from the crater B floor subsequent to the fluvial and lacustrine activity. Insight into the nature of depositional and inversion processes can be gained from analyses of the larger and more extensive inverted channels associated with regional valley networks. Davis et al. (2016, 2019) characterized large inverted fluvial channels in Arabia Terra (generally 1-2 km wide, >10 km long). Most of these are found immediately downslope of, or deposited within, regionally integrated valley networks, or within the same watershed if not directly associated with a valley network. A global survey by Dickson et al. (2020) reached similar conclusions about the distribution of large inverted fluvial channels (>70 m wide), finding that all were downslope of valley networks. These types of inverted channels most likely represent the depositional counterparts to traditional valley networks forming in the Late Noachian—Early Hesperian (Davis et al. 2016, 2019; Dickson et al. 2020).

In the case of inverted fluvial channels in Arabia Terra, Davis et al. (2016, 2019) noted that valley to ridge transitions are correlated with slope breaks, indicating that the ridges are likely to have formed out of alluvial channel deposits. Fassett and Head (2007) identified inverted channels in Arabia Terra in association with a Noachian—Hesperian mantling deposit. The characteristics of this mantling unit suggest that volatiles were incorporated either during or immediately following its emplacement. They suggested that the inversion process involved sublimation and volatile loss, combined with aeolian deflation of unconsolidated material, leaving behind indurated fluvial channel deposits.

Topographic data indicate that the inverted channels in crater B initiate at a slope break at the base of the crater wall (Figs. 2A, 3), which matches the depositional style inferred by Davis et al. (2016, 2019). The thermophysical properties of the channels from THEMIS quantitative thermal inertia data (Fergason et al. 2006) are intermediate between the high-TI, light-toned crater floor bedrock and the relatively low-TI mantling materials on the crater rim and wall. According to Williams et al. (2018), this would suggest that the inverted channels are more competent than the surrounding loose sediment and were formed through induration. However, spectral identifications of primary mafic mineral phases in the inverted channels are consistent with limited post-depositional alteration (Irwin et al. 2018), and HiRISE images of the channels show active weathering and talus slopes uncharacteristic of indurated materials, so grain armoring cannot be ruled out as an alternative.

We interpret the fluvial channel deposits in crater B to have originally contained alluvial sediment lags derived from mantling deposits within the crater. The nature of the mantling deposit is likely to have been similar to those observed in Arabia Terra, and elsewhere in Terra Sabaea, consisting of airfall deposits from climate-driven dust and volatile deposition, atmospherically transported impact ejecta, or plinian volcanic tephra (Fassett & Head 2007). Fluvial processes concentrated sediments within the fluvial channels and transported suspended loads either into lakes or floodplains, where they were eventually removed or redistributed. Topographic inversion resulted from aeolian deflation, perhaps combined with sublimation of co-deposited volatile layers and cement, as suggested by Fassett and Head (2007) for the mantling unit in Arabia Terra. The source of the alluvial sediment remaining in the inverted channels is thus likely to have been from previous mantle emplacement and/or material shed from the adjacent crater wall.

## 3. WATER SOURCES IN THE CSDB CRATER

The ridges and basins documented within the CSDB crater B (Figs. 1–3) indicate an integrated fluvial–lacustrine system that formed completely internally to the crater basin with no apparent breaches or other incoming fluvial drainage sources or outgoing drainage sinks. Fluvial channels within the crater subsequently underwent topographic inversion into ridges, most likely involving aeolian removal of loosely consolidated sediment and/or sublimation of volatiles. We compare our observations in crater B to previously presented models of crater degradation and fluvial incision on Mars to determine the most plausible water sources in the crater and the origins of the ridges, basins, alcoves, and upslope-facing scarps.

### 3.1. Regional Fluvial Processes

The distribution of degraded Noachian-aged craters in the southern highlands suggests that the processes responsible for their degradation were widespread (Craddock & Maxwell 1993; Craddock et al. 1997; Irwin et al. 2013). Dawes crater (Fig. 4B) has many of the hallmarks described by Craddock and Howard (2002) and others as typical of Noachian crater degradation: the crater wall has been heavily incised and backwasted by fluvial erosion, and the flat, relatively featureless floor has been infilled presumably by laterally transported sediments. By comparison, the wall and rim of crater B (Fig. 4A) are much more subdued, and no fluvial incision is observed. The floor of crater B contains a dense network of inverted fluvial channels that begin abruptly at the base of the crater wall and do not extend up to or beyond the rim. For these reasons, we believe external drainage from a regionally integrated valley network or inlet channels is the least likely water source in crater B.

### 3.2. Groundwater Flow Processes

Irwin et al. (2018) noted the possibility of groundwater flow into crater B due to higher surrounding intercrater topography. Many groundwater-fed crater basin lakes elsewhere on Mars contain steep, short valleys that erode into the crater wall. Salese et al. (2019) interpreted these as groundwater sapping valleys, noting their low drainage density and approximately equal channel width downstream. Crater B shows no evidence for groundwater sapping valleys as described by Salese et al. (2019), and the higher drainage density and stream order of the inverted channels on the crater floor are atypical of groundwater sapping processes. While a few groundwater-fed lakes do display inverted fluvial morphologies, these are almost always associated with sapping valleys (Salese et al. 2019). Although groundwater-fed lakes appear to be less filled with sediment than other crater basin lakes (Salese et al. 2019), the fluvial activity resulting from groundwater sapping was still sufficient to transport sediment into fans and deltas (Fig. 4C). Crater B contains no discernible landforms of this nature. The upslope-facing scarps appear superficially similar to some features found in certain groundwater-fed crater basin lakes; however, in crater B, the steeper sides face upslope (Figs. 2B-C), in the opposite sense of a typical terrace that, in a groundwater-fed lacustrine environment, would be downcut on the side facing downslope. Finally, the crater B floor elevation (+2 km) is inconsistent with the elevation that groundwater-fed lakes on Mars occupy, usually below −4 km (Salese et al. 2019). Thus, we believe that groundwater sapping is an unlikely water source in crater B.

### 3.3. Glacial Meltwater Processes

Thus far, we find that our observations in crater B do not match craters modified by rainfall-derived fluvial breaching or groundwater sapping. As noted by Craddock and Howard (2002), "…it is possible that…any surface runoff was the result of snowmelt." Snowmelt may refer to any kind of melting resulting from the heating of surficial water in its solid phase, and we thus investigate this option as a water source in crater B.

Could glaciation have occurred in the highlands of Noachian Mars, forming a readily available water source for the crater floor fluvial and lacustrine features in crater B? Such a scenario is one potential outcome of the colder climate predicted by early Mars climate modeling studies (e.g. Forget et al. 2013; Wordsworth et al. 2013, 2015) and is consistent with typical Noachian crater wall slopes (Kreslavsky & Head 2018). Model-based predictions of the likely range of geothermal heat fluxes and surface temperatures in the Noachian further suggest that any glacial activity would have been cold-based (Fastook & Head 2015), generally ruling out the formation of basal melt features such as eskers that can appear superficially similar to inverted fluvial channels (e.g. Butcher et al. 2021). We further disfavor an esker interpretation for the ridges in crater B because 1) eskers are often observed flowing along or against gradient, while the ridges in crater B flow strictly along gradient; and 2) eskers may have a range of ridge crest morphologies including sharp peaks, while the ridges in crater B are almost exclusively flat-topped (Butcher et al. 2021).

Instead, top-down (supraglacial) melting of a cold-based glacier could occur in cases where air temperatures locally exceeded 273 K (e.g. Head & Marchant 2014). In the more recent Amazonian, cold-based glaciation in crater interiors has produced a variety of distinctive morphological features. Ice deposits on steep slopes underwent glacial flow (Fastook & Head 2014) down to the exposed base of crater walls, forming terminal and lateral moraines whose boundaries are preserved as upslope-facing scarps (Berman et al. 2005; Jawin et al. 2018; Fig. 4F). In larger craters with shallower wall slopes, numerous glaciofluvial valley networks formed as a result of glacial meltwater drainage below the scarps (Berman et al. 2009; Fassett et al. 2010).

Much as in the more recent Amazonian examples, glacial ice cold-trapped in the alcoves of crater B would have undergone flow, ablation, and melting, resulting in the buildup of marginal sediments at the base of the crater wall that now comprise the upslope-facing scarps and their slope-parallel extensions. Additional top-down melting at the glacial front would then result in proglacial fluvial erosion of sediments derived either directly from glacial debris deposition (e.g. Head et al. 2017; Denton & Head 2017) or from preexisting aeolian mantles (Fassett & Head 2007). Proglacial meltwater channels are also observed in conjunction with seasonal top-down melting processes in the Antarctic McMurdo Dry Valleys (e.g. Atkins & Dickinson 2007; Atkins 2013; Head & Marchant 2014). In some cases, ridges may form between active channels (Atkins & Dickinson 2007; Atkins 2013) as the result of stranded ice-cored ground exposed by glacial retreat; this process is distinct from the topographic inversion that we interpret for the formation of the ridges in crater B.

We conclude that episodic top-down melting of a cold-based crater wall glacier within crater B represents a plausible water source to form the observed fluvial and lacustrine features. This CSDB water source differs from previously considered hypotheses for the formation of other types of paleolakes on Mars, which call on water sources external to the crater (Fassett & Head 2008b; Goudge et al. 2015; Salese et al. 2019). In addition, ambient cold climates would favor the emplacement of climate-driven snow, ice and dust mantles (e.g. Head et al. 2003). The loss of such volatile components by ablation could significantly assist in the topographic inversion and mantle loss processes in crater B, as also envisioned in Arabia Terra (Fassett & Head 2007).

## 4. CONCLUSIONS

We describe a Noachian closed-source drainage basin (CSDB), a new type of paleolake on Mars with neither inlet nor outlet channels. An ensemble of ridges, basins, alcoves, and upslope-facing scarps found within the crater provides evidence for possible drainage sources and mechanisms leading to the formation of inverted fluvial channels and lacustrine deposits on the crater floor (Figs. 1-3). We find that the features within the crater are unlikely to have formed in a warm and wet early Mars climate dominated by fluvial valley network incision and crater breaching. The occurrence of inverted channels within the crater differs from descriptions of large, regionally integrated inverted channels that appear to be the depositional counterparts to traditional valley networks (Davis et al. 2016, 2019; Dickson et al. 2020). The crater also does not share many of the characteristics of groundwater-fed crater basin lakes (fans, deltas, shorelines; Salese et al. 2019), and we thus consider groundwater to be an unlikely source of water in the CSDB crater.

Alternatively, we have explored the suggestion of Craddock and Howard (2002) that surface runoff and crater degradation may have been the result of snowmelt. We find that the ensemble of features within the crater (e.g. lack of wall-breaching fluvial channels, the presence of alcoves and upslope-facing scarps, and the nature and location of inverted fluvial channels and lacustrine deposits on the crater floor) are all consistent with the top-down melting of a cold-based crater wall glacier, proglacial drainage in fluvial meltwater channels, and ponding of meltwater in lakes on the crater floor.

The interpretation of the features in the CSDB crater as the result of top-down melting of a cold-based crater wall glacier in the southern highlands has significant implications for our understanding of the ambient early Mars climate. Several workers have described mechanisms for transient atmospheric heating and ice melting in an otherwise subfreezing climate (Wordsworth et al. 2015; Fastook & Head 2015; Palumbo et al. 2018), but specific geomorphic evidence for Noachian glaciofluvial activity has only occasionally been suggested (e.g. Craddock & Howard 2002; Bouquety et al. 2019; Butcher et al. 2021). Here, we describe for the first time observational evidence of proglacial paleolakes and inverted fluvial channels on Noachian Mars.

The proglacial fluvial and lacustrine features in this closed-source drainage basin provide the necessary criteria for recognizing distinctive geomorphic features that could be formed via cold-based glacial processes elsewhere on Mars. Our initial investigations of the broader region surrounding crater B (Boatwright & Head 2021) indicate that these types of features are widespread within craters of similar age, particularly along the high-standing escarpments of the Hellas basin rim. Among these additional examples, crater B appears to represent an end member with the most well-preserved examples of proglacial fluvial and lacustrine features within a CSDB crater. This ensemble of features can be compared to those in other nearby craters in order to place crater B into a regional and global geologic context and to assess both ambient and changing climate conditions (e.g. Ramirez & Craddock 2018; Wordsworth et al. 2015; Palumbo et al. 2018).

ACKNOWLEDGEMENTS

We gratefully acknowledge support to JWH for participation in the Mars Express High Resolution Stereo Camera Team (JPL 1237163). We thank Jay Dickson and one anonymous reviewer for helping substantially improve the manuscript. The authors declare no competing financial interests. Datasets generated for this paper are archived in the Harvard Dataverse at https://doi.org/10.7910/DVN/QSO2PJ.ACKNOWLEDGEMENTS

We gratefully acknowledge support to JWH for participation in the Mars Express High Resolution Stereo Camera Team (JPL 1237163). We thank Jay Dickson and one anonymous reviewer for helping substantially improve the manuscript. The authors declare no competing financial interests. Datasets generated for this paper are archived in the Harvard Dataverse at https://doi.org/10.7910/DVN/QSO2PJ.

## APPENDIX A. IMAGE DATA SOURCES

Individual images and mosaics used in the figures are listed in Table A1. The CTX stereo DEM was generated with the Ames Stereo Pipeline (Beyer et al. 2018). Topographic profiles were measured from the DEM in ArcMap using the Interpolate Line tool. The global CTX image mosaic is described by Dickson et al. (2018).

| Figure | Instrument | Image |
|---|---|---|
| 1 | CTX | B16_015983_1596_XI_20S317W<br>P12_005540_1595_XI_20S_317W<br>P14_006608_1596_XI_20S317W<br>CTX_Mosaic_Beta01_Feb2018 |
| 2 | | B10_013649_1596_XI_20S317W<br>B16_015983_1596_XI_20S317W<br>25 m/pix stereo DEM:<br>  B10_013649_1596_XI_20S317W<br>  B11_013794_1596_XI_20S317W<br>  B16_015983_1596_XI_20S317W<br>  B17_016128_1595_XI_20S317W |
| 3 | | B16_015983_1596_XI_20S317W<br>25 m/pix stereo DEM (same images) |
| 4A | | B16_015983_1596_XI_20S317W<br>P12_005540_1595_XI_20S_317W<br>P14_006608_1596_XI_20S317W<br>CTX_Mosaic_Beta01_Feb2018 |
| 4B | THEMIS | THEMIS-IR-Day_mosaic_global_100m_v12 |
| 4C, E-F | CTX | CTX_Mosaic_Beta01_Feb2018 |
| 4D | | B16_015983_1596_XI_20S317W<br>P14_006608_1596_XI_20S317W |

Table A1. Image data used for figures in the main text.

## APPENDIX B. PALEOHYDROLOGIC RECONSTRUCTION METHODS

Hayden et al. (2019) used the following equation to calculate bankfull river discharge:

$$Q = 74.2 C_f^{-1/2} (Rg)^{3/8} v^{1/4} D_{50}^{1/8} d^2 \tag{B1}$$

Where $C_f$ is the coefficient of friction, $R$ is the submerged specific gravity of sediment, $g$ is the gravitational acceleration, $v$ is the kinematic viscosity of water, $D_{50}$ is the median grain size, and $d$ is the paleochannel depth. Hayden et al. (2019) specifically noted that the measured caprock thickness is likely to be ~1.5x the original paleochannel depth $d$. They also noted that the current slope of the geomorphic surface is likely not the original bed slope; instead, they inferred the bed slope $S$ from the bankfull Shields stress $\tau^*$ expressed as a function of the particle Reynolds number $Re_p$:

$$S = R D_{50} \tau^* / d \tag{B2}$$

$$\tau^* = 17 Re_p^{1/2} \tag{B3}$$

$$Re_p = (R g D_{50})^{1/2} D_{50} / v \tag{B4}$$

Stream velocity $U$ was calculated by Hayden et al. (2019) as:

$$U = \left(\frac{1}{\kappa}\right) \ln\left[11 \left(\frac{R D_{50}}{S k_s}\right) \tau_s^*\right] (R g D_{50} \tau_s^*)^{1/2} \tag{B5}$$

Where $\kappa$ is von Karman's constant, $k_s$ is the grain roughness length scale, and $\tau_s^*$ is the skin friction component of the Shields stress:

$$\tau_s^* = 0.06 + 0.4 \tau^{*2} \tag{B6}$$

Caprocks are not distinguishable at the scale of our DEM, so for channel depth $d$, we conservatively assume a caprock thickness that is 10% the total ridge height, and use the minimum and maximum measured ridge heights for the inverted channels in crater B to bracket a possible range of $d$ = 0.7–1.3 m. We use a median grain size of medium-coarse sand, $D_{50}$ = 0.5 mm.

For the above values, we find stream velocities $U$ = 2.5–2.7 m/s and bankfull discharges $Q$ = 2,000–4,400 m³/s per channel for the trunk streams that flow into the basins.

We then calculate the total time that would be required to fill the two basins in crater B with a sediment cover 50 m thick, which is approximately the difference in elevation between the deepest part of the basins and the channel termination points. This results in a combined sediment volume of ~1.7x10¹⁰ m³.

In order to determine the amount of sediment that flowed through the channels for a given discharge, we use the fluid-sediment ratio relation of Rosenberg et al. (2019):

$$Q_s = \frac{1.5 T^{3/2} D_{50}}{d} \left(\frac{v^2}{D_{50}^3 gR}\right)^{1/10} Q_f \tag{B7}$$

Where $T$ is a transport stage parameter for the initiation of grain motion, taken to be ~91 for the channels in crater B. If we assume that only channel was active at a time, the sediment flux into the basins could range from $Q_s$ = 450–490 m³/s. Based upon these results, we find that the basins could have been filled to a depth of 50 m in ~1 Earth year of constant flow.

It is more likely, however, that flow in the channels was intermittent. If we assume one day of flow per martian year during peak summertime heating (e.g. Palumbo et al. 2018), this timescale can be extended perhaps by two orders of magnitude. Regardless, this is significantly shorter than the timescale believed to be required to form the valley networks, or ~$10^5$–$10^7$ years (e.g. Hoke et al. 2011).

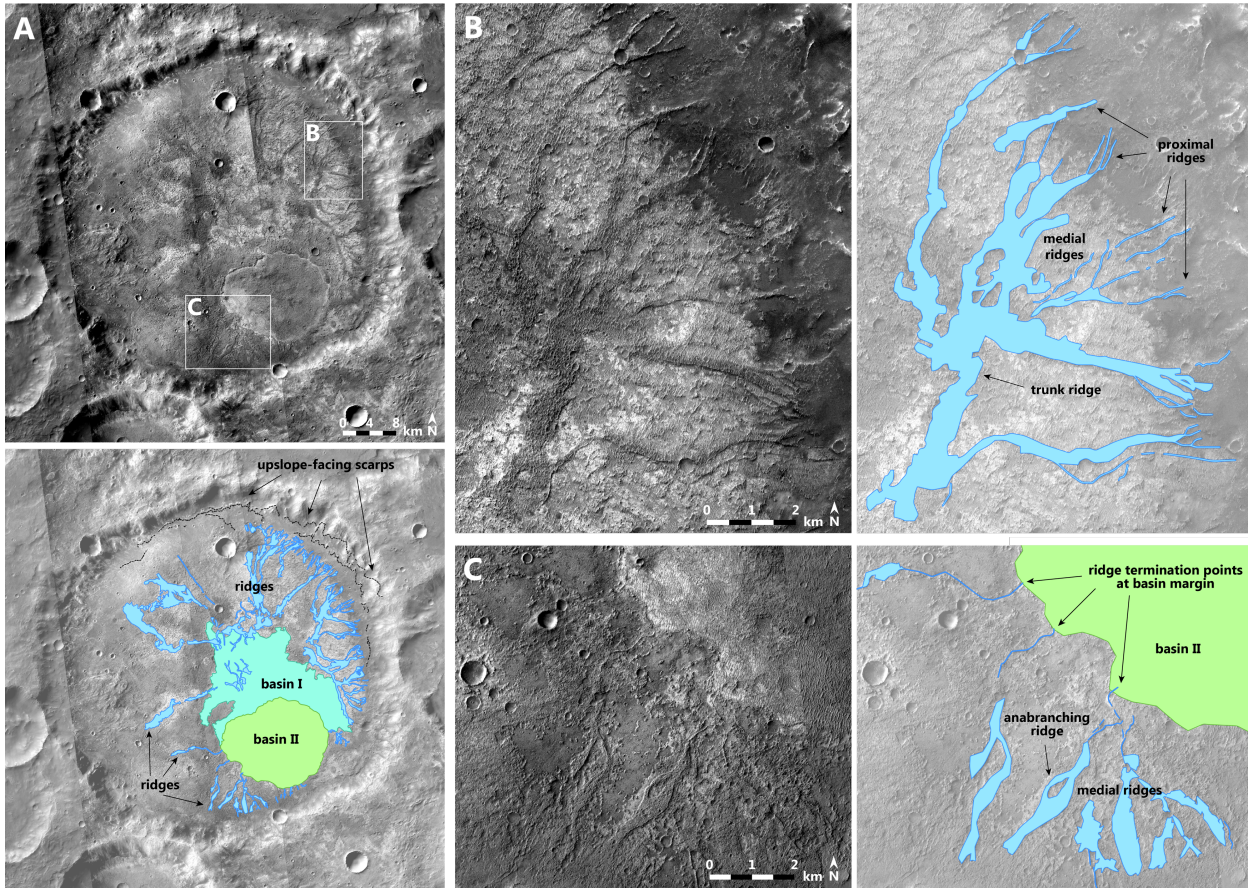

Fig. 1. Visible images with sketch maps showing major geologic features in crater B. (A) The 54-km diameter CSDB crater "B" (20.3°S, 42.6°E). No inlet or outlet breaches are observed. Ridge networks (blue) begin near the crater wall base, where they are coincident with upslope-facing scarps (dashed lines) and crater wall alcoves. Ridges terminate in two paleolake basins (cyan and green). (B) The morphology of the eastern ridge network transitions downslope from dense proximal ridges to more widely spaced medial ridges; several medial ridges then converge into a single trunk ridge. (C) The southern ridge network terminates at three separate points along the margin of basin II. One of the ridges also displays an anabranching morphology.

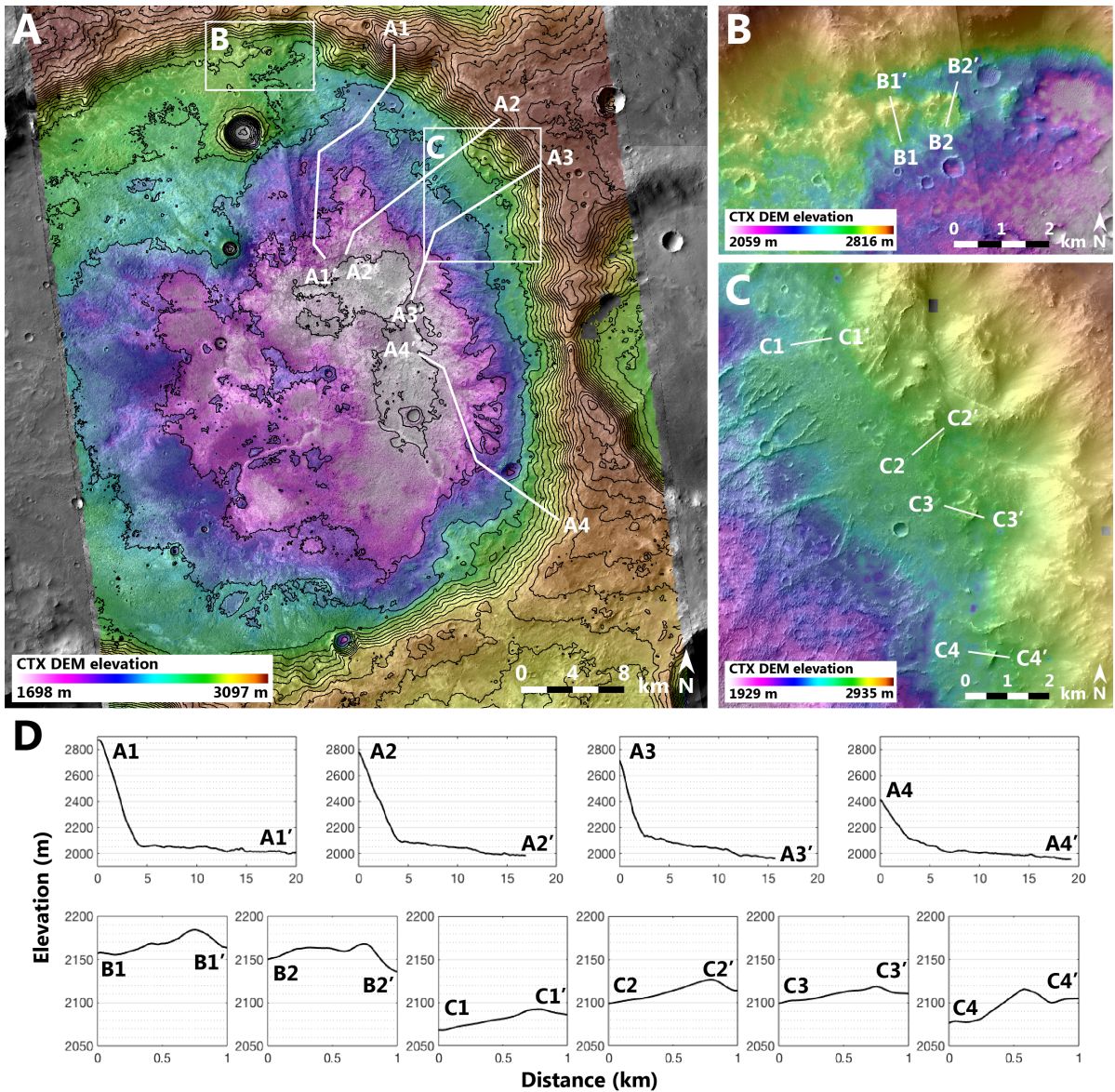

Fig. 2. Topography of crater B. (A) Topographic data confirms the tributary nature of the ridge networks and relative positions of the two depositional basins. White lines are ridge topographic profiles A1–A4; black lines are 50 m contours; white boxes show locations of insets B–C. (B–C) Close-up views of upslope-facing scarps in the north and east of the crater with locations of topographic profiles marked (white lines B1–B2 and C1–C4). (D) Topographic profiles of four ridges with their respective upslope crater wall segments (top) and six well preserved scarp segments (bottom). Each profile clearly displays a slope break at the base of the crater wall where proximal ridges initiate. Each scarp segment is asymmetric (steeper side upslope) and displays a similar morphology and degree of relief above the floor of ~10–30 m.

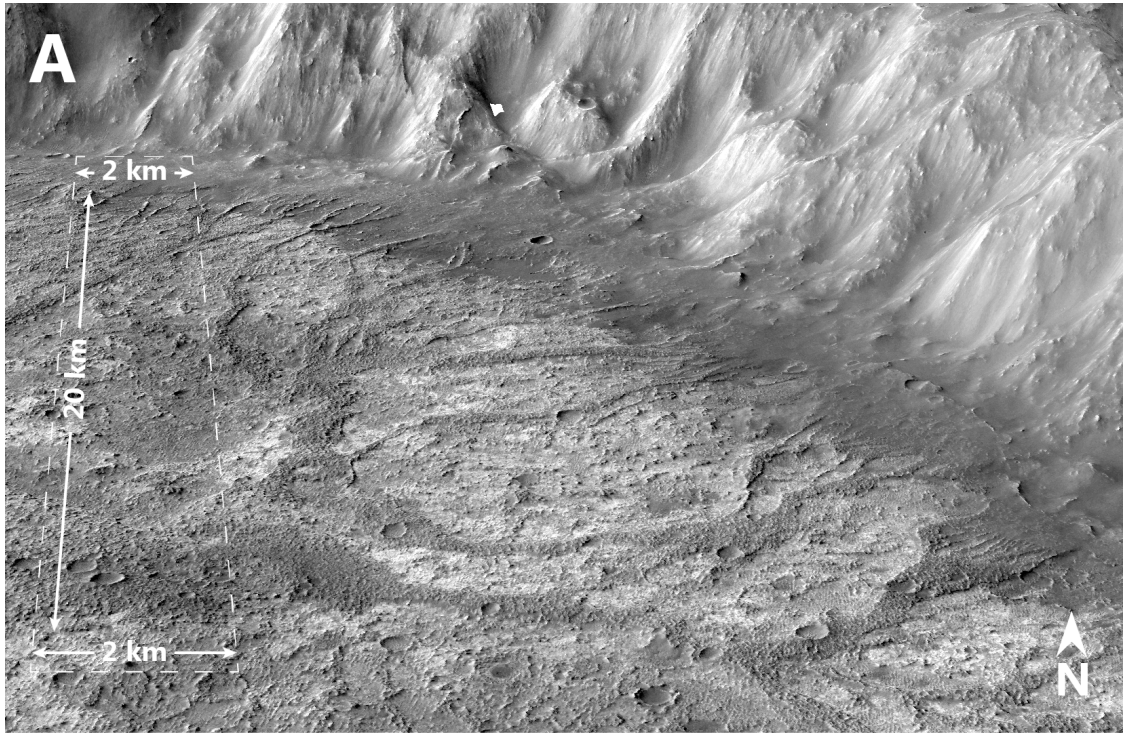
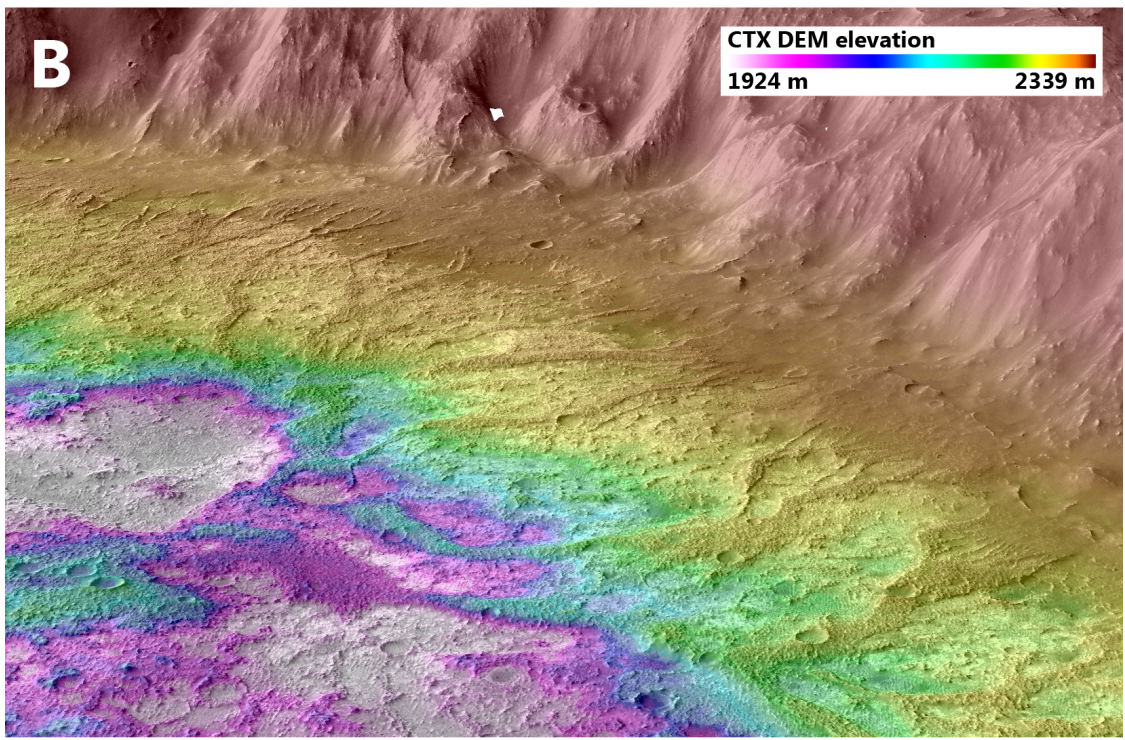

Fig. 3. Three-dimensional perspective view of crater B. Upslope-facing scarps and proximal ridge heads occur at a slope break below the crater wall base. Toward the crater interior, the positive relief of the wider medial ridges against the rugged floor is apparent. 4x vertical exaggeration.

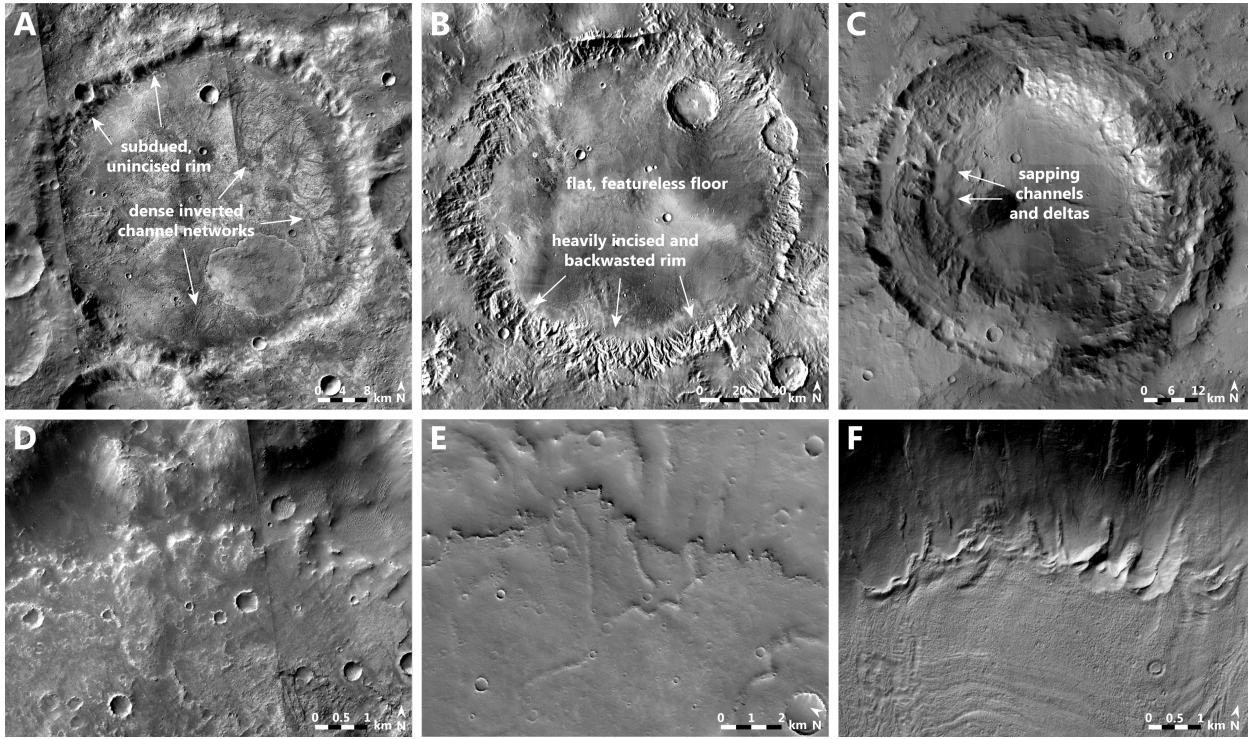

Fig. 4. Comparison of crater B to typical degraded craters on Mars. (A) Defining features of crater B include a subdued, unincised rim and dense crater floor inverted channel networks. (B) Dawes, a "classic" degraded crater (9.2°S, 38.0°E), has a heavily incised and backwasted rim with a flat, relatively featureless floor. (C) A typical groundwater-fed crater (27.9°N, 11.5°E) displays sapping channels and deltas. The three craters are shown at the same relative scale in order to facilitate comparison. (D) Laterally continuous segment of upslope-facing scarps in the northern floor of crater B. (E) Upslope-facing scarp identified by Davis et al. (*23*) in Barth crater in Arabia Terra (7.4°N, 25.7°E). (F) Upslope-facing scarps formed by retreating cold-based glaciers in an Amazonian-aged crater in Terra Sirenum, southwest of Newton basin (42.9°S, 196.7°E).